\documentclass[prl,aps,reprint,preprintnumbers,amsmath,amssymb,superscriptaddress]{revtex4-1}

\usepackage{dsfont}
\usepackage{graphicx}
\usepackage{dcolumn}
\usepackage{footmisc}
\usepackage{setspace}
\usepackage{caption}
\DeclareMathOperator{\arctanh}{arctanh}

\newcommand\fft[2]{\frac{#1}{#2}}
\newcommand\nn{\nonumber}

\usepackage{xcolor}
\definecolor{plotorange}{RGB}{216,138,28}
\definecolor{plotblue}{RGB}{76,108,166}

\begin{document}

\preprint{LCTP-20-27}

\title{A NUT Charge Weak Gravity Conjecture from Dimensional Reduction}
\author{Sera Cremonini}
\email{cremonini@lehigh.edu}
\affiliation{
Department of Physics, Lehigh University, Bethlehem, PA, 18018, USA
}
\author{Callum R. T. Jones}
\email{cjones@physics.ucla.edu}
\affiliation{
Mani L. Bhaumik Institute for Theoretical Physics, Department of Physics and Astronomy, University of California Los Angeles, Los Angeles, CA 90095, USA
}
\affiliation{Leinweber Center for Theoretical Physics, Randall Laboratory of Physics, University of Michigan, Ann Arbor, MI 48109-1040, USA}
\author{James T. Liu}
\email{jimliu@umich.edu}
\affiliation{Leinweber Center for Theoretical Physics, Randall Laboratory of Physics, University of Michigan, Ann Arbor, MI 48109-1040, USA}
\author{Brian McPeak}
\email{bmcpeak@umich.edu}
\affiliation{Leinweber Center for Theoretical Physics, Randall Laboratory of Physics, University of Michigan, Ann Arbor, MI 48109-1040, USA}
\affiliation{Department of Physics, University of Pisa, Largo B. Pontecorvo 3, 56127 Pisa, Italy}
\author{Yuezhang Tang}
\email{yut318@lehigh.edu}
\affiliation{
Department of Physics, Lehigh University, Bethlehem, PA, 18018, USA
}

\begin{abstract}
    We analyze the constraints on four-derivative corrections to 5d Einstein-Maxwell theory from the black hole Weak Gravity Conjecture (WGC). We calculate the leading corrections to the extremal mass of asymptotically flat 5d charged solutions as well as 4d Kaluza-Klein compactifications. The WGC bounds from the latter, interpreted as 4d dyonic black holes, are found to be strictly stronger. As magnetic graviphoton charge lifts to a NUT-like charge in 5d, we argue that the logic of the WGC should apply to these topological charges as well and leads to new constraints on purely gravitational theories.
\end{abstract}

\maketitle

\section{Introduction}

Black holes have been a fruitful source of insight into many questions about the nature of quantum gravity. One such question is whether any low-energy effective field theory (EFT) containing semiclassical Einstein gravity coupled to matter, assumed to be free from any IR pathologies, can be UV completed to a full quantum theory of gravity. The answer appears to be negative \cite{Vafa:2005ui}, and theories which do not have any such UV completion are said to lie in the \textit{Swampland}.

One proposed criterion to demarcate the Swampland is the \textit{Weak Gravity Conjecture} (WGC). This conjecture asserts that, for each rational direction in charge space, there should not exist an infinite tower of exactly stable, non-BPS, extremal black holes. Originally motivated by the desire to avoid having an unbounded number of stable states below a given mass scale as the gauge coupling is tuned to zero \cite{ArkaniHamed:2006dz}, and the associated problems with remnants \cite{Susskind:1995da},
the necessity of the WGC has subsequently been argued from the covariant entropy bound \cite{Banks:2006mm}, the modular bootstrap \cite{Benjamin:2016fhe,Montero:2016tif}, weak cosmic censorship \cite{Crisford:2017gsb} and beyond. For a recent review of the WGC and the Swampland program more generally, see \cite{Palti:2019pca}.

The stability of large, asymptotically flat, extremal black holes depends on the spectrum of charged states in the theory. The decay is kinematically allowed only if there exists some object among the decay products with charge-to-mass ratio exceeding the extremality bound of the parent black hole \cite{ArkaniHamed:2006dz}. This may be an elementary particle, but can also be a smaller extremal black hole if higher-derivative or quantum corrections \textit{decrease} the extremal mass at fixed charge \cite{Kats:2006xp}. The possibility that the latter resolution, with the associated bounds on the Wilson coefficients controlling the mass shift, is \textit{always} realized for a model in the Landscape, corresponds to a stronger or refined ``black hole"  version of the WGC.

Some evidence in favor of this refined conjecture comes from the fact that, in many cases, it is possible to prove the required bounds using unitarity, causality, RG flow and the relation with the entropy shift \cite{Cheung:2014vva, Cottrell:2016bty, Chen:2019qvr, Cheung:2018cwt, Bellazzini:2019xts, Andriolo:2018lvp, Hamada:2018dde, Charles:2019qqt, Jones:2019nev, Goon:2019faz}. The refined conjecture has also been extended to a broader set of models, including multiple gauge fields and/or scalar fields, a cosmological constant, and settings with additional duality symmetries \cite{Jones:2019nev, Cremonini:2019wdk,Loges:2019jzs,Andriolo:2020lul}.

In this letter we argue that the WGC should be extended to include topological charges associated with the global structure of spacetime \cite{Hull:1997kt}. Specifically, the topological charge associated to the Taub-NUT factor of Kaluza-Klein (KK) monopoles must be treated on equal footing with gauge charges to produce a set of WGC bounds that are self-consistent under dimensional reduction. A general, $d$-dimensional, definition of these topological charges is discussed in detail in \cite{Hull:1997kt}. Higher dimensional generalizations of the KK monopole have been discussed in many contexts \cite{Lee:1984ica,Mann:2005gk}, and play an important role \emph{e.g.} in the M-theory/type IIA correspondence \cite{Sen:1997js}. In 5d, such configurations are possible even without primordial gauge fields and therefore lead to WGC constraints on low-energy EFTs of pure gravity. For models which do contain primordial gauge fields, considering combinations of gauge and topological charges leads to stronger bounds than either alone.

We illustrate these claims in the context of 5d Einstein-Maxwell theory. Beginning with a general parametrization of four-derivative corrections to the effective action, we determine the WGC bounds arising from 5d, asymptotically flat, extremal, electric black hole and magnetic black string solutions. The latter appear not to have been discussed in the literature so far. We then consider solutions with one of the dimensions compactified on $S^1$. In the 4d KK reduction of the EFT, we study a set of two-charge solutions and calculate the associated WGC bounds. Comparing the 4d and 5d bounds, we find a set of solutions -- specifically, those which are magnetically charged under the graviphoton in the 4d description -- which strengthen the bounds relative to those implied by the 5d solutions alone. In the 5d uplift, these solutions correspond to charged KK monopoles with the 4d magnetic graviphoton charge corresponding to a NUT charge in the 5d spacetime \cite{Gross:1983hb,Sorkin:1983ns}. Demanding that the set of WGC bounds is consistent in both the 4d and 5d description requires enlarging the statement of the 5d WGC to include such topological charges.

In this paper we assume that the dominant correction to the black hole extremality bound is given by the classical (tree-level) contribution from four-derivative operators in the low-energy effective action, and in particular that quantum (one-loop) corrections are sub-dominant. In $d>4$ this is a trivial consequence of the fact that four-derivative operators are not renormalized due to the dimensionality of Newton's constant. The validity of our analysis after dimensional reduction to $d=4$, meaning the sub-dominance of 4d quantum corrections, requires that we assume that the KK mass scale $M_{\text{KK}}$ is smaller than a certain combination
\begin{equation}
    M_{\text{KK}}\ll\frac{ M_5^3}{\Lambda^{2}},
\end{equation}
where $\Lambda$ is the scale of the physics generating the four-derivative operators and $M_5$ is the 5d Planck scale. For the remainder of this paper we will assume this hierarchy is satisfied. In principle, we should also consider the effect of integrating out KK modes. However, our 4d theory is a consistent truncation of the 5d theory, so these cannot couple at tree-level. Thus, the above discussion on 4d quantum loops applies to this case as well. 

Our setup is similar to \cite{Heidenreich:2015nta}, which also found that the WGC is strengthened by dimensional reduction through the appearance of additional KK gauge fields. That work primarily concerned constraints on the spectrum of elementary charged particles, and did not analyze the role of graviphoton magnetic charges (or of higher derivative corrections). In this letter we use a similar setup to address the complementary problem of constraining the spectrum of extremal black holes and the associated Wilson coefficients appearing in the low-energy EFT.

Finally, we discuss how the bounds are affected by supersymmetry, which relates the EFT coefficients to each other. Indeed, if we require that the four-derivative action of the 5d theory has $\mathcal{N} = 2$ supersymmetry, there will be only one independent EFT coefficient. We show that the extremal mass shift vanishes for all the 5d charged black backgrounds, but not for the topological background discussed above. Thus, our WGC bound for topological charges implies a bound on the sign of the EFT coefficients even in the supersymmetric theory, where the previous forms of the WGC did not apply.

While preparing this work, we learned about \cite{NewSpin}, which provides a different argument for the strengthening of WGC bounds under dimensional reduction.

\section{Constraints from 5D Black Objects}
\label{sec:5DWGC}

We are interested in 5d theories described, at leading order in derivatives, by the effective Lagrangian
\begin{align}
-16\pi \hat{e}^{-1}\mathcal{L}_5=
    \hat{R} - \frac{1}{4} \hat{F}^2 + \lambda
    \epsilon^{\mu\nu\rho\sigma\lambda}\hat{A}_\mu\hat{F}_{\nu\rho}\hat{F}_{\sigma\lambda}\, ,
    \label{eq:2derLag}
\end{align}
where hats denote 5d quantities, and we have included a Chern-Simons term for completeness. Pure Einstein-Maxwell theory corresponds to $\lambda=0$, while the bosonic sector of $\mathcal{N} = 2$ supergravity in 5d corresponds to $\lambda^{-1}=12\sqrt3$. however, in this letter we only consider black hole geometries that are insensitive to $\lambda$.

The black hole WGC asserts that, when higher-derivative operators are included in the effective action, the mass of an extremal solution is \textit{decreased} for a given charge. A complete basis (up to field redefinitions) of the leading four-derivative operators in 5d Einstein-Maxwell theory is given by
\begin{eqnarray}
\label{finalL}
-16\pi \hat{e}^{-1} \, \Delta \mathcal{L}_5 &=&
c_1 \hat{R}_{GB} +
c_2 \, \hat{W}_{\mu\nu\rho\lambda} \hat{F}^{\mu\nu} \hat{F}^{\rho\lambda}  + c_3 (\hat{F}^2)^2~~
\nonumber \\
 &&  
 + \, c_4 \hat{F}^4  +  c_5 \, \epsilon^{\mu\nu\rho\lambda\sigma} \hat{A}_\mu \hat{R}_{\nu\rho\delta\gamma} \hat{R}_{\lambda \sigma}^{\;\;\;\; \delta \gamma}  
 \,, 
\end{eqnarray}
where $\hat{R}_{GB}=\hat{R}_{\mu \nu \rho \sigma} \hat{R}^{\mu \nu \rho \sigma} - 4 \hat{R}_{\mu \nu} \hat{R}^{\mu \nu} + \hat{R}^2$ is the Gauss-Bonnet combination, $\hat{W}$ is the Weyl tensor, $ (\hat{F}^2)^2 = (\hat{F}_{\mu \nu} \hat{F}^{\mu \nu})^2$ and $\hat{F}^4 = \hat{F}_{\mu \nu} \hat{F}^{\nu \rho} \hat{F}_{\rho \sigma} \hat{F}^{\sigma \mu}$. 
%
%

For $\mathcal{N} = 2$ supergravity, the coefficients $c_i$ are related by supersymmetry. In particular, the supersymmetric completion of $\hat A\wedge\hat R\wedge\hat R$ in gauged supergravity was computed in  \cite{Hanaki:2006pj,Cremonini:2008tw}. After changing to the same basis of operators we use here and taking the ungauged limit, we find that the Wilson coefficients $c_i$ can be expressed in terms of a single coefficient $c$ \footnote{In the gauged theory, the parameter $c$ can be related to the central charges of the dual CFT using anomalies \cite{Cremonini:2008tw}.} as 
\begin{align}
\label{SUSYcoeff}
    c_1 = -2 \, c_2 = -6 \, c_3 =  \frac{24}{11} \, c_4 = 2\sqrt{3} \, c_5 \equiv c.
\end{align}
For completeness in (\ref{finalL}) we have allowed for the gauge gravitational Chern-Simons term (proportional to $c_5$), but this operator gives a vanishing correction to the solutions considered in this letter, hence we do not constrain it in our present analysis.

The $\mathcal{O}(c_i)$ correction to thermodynamic quantities, including the extremal mass, may be obtained from the Helmholtz free-energy $F$, which is related to the on-shell Euclidean action $I_E$ as $\beta F(Q,T) = I_E(\hat{g}_E,\hat{A}_E)$. Computing the four-derivative corrections to the on-shell action does \textit{not} require that we find the corresponding corrections to the geometry; it was shown in \cite{Reall:2019sah} that it is sufficient to evaluate the four-derivative operators on the two-derivative backgrounds. See \cite{Cremonini:2019wdk} for a detailed discussion of how to calculate the on-shell action and obtain the mass shift, including the details of the regularization procedure and inclusion of appropriate boundary terms.

The black hole WGC asserts that for all extremal black objects, $(\Delta M)_{T=0}<0$. This is possible only if the Wilson coefficients $c_i$ satisfy certain constraints. Below we calculate these constraints for 5d asymptotically flat black holes and strings
\footnote{Since the gravity solutions we consider are standard, we only summarize their salient features.  The explicit form of the solutions and additional thermodynamic quantities are provided in the Appendix.}.\\
\\
\noindent
{\bf \underline{5d Electric Black Holes}:}
We begin with the familiar 5d Reissner-Nordstrom black holes that are characterized by their mass $M$ and electric charge $Q$, subject to the two-derivative extremality bound
\begin{equation}
    M \geq |Q|.
\end{equation}
Extremal black holes saturate this bound and have zero temperature. By computing the four-derivative correction to the Helmholtz free-energy, one finds the mass shift at extremality
\begin{align}
\label{5DshiftE}
    (\Delta M)_{T = 0} = -\frac{3 \pi}{8} (-2 c_1  -c_2 + 24 c_3 +12 c_4) \, .
\end{align}
Up to a choice of EFT basis this expression agrees with the $d$-dimensional result calculated in \cite{Kats:2006xp}.  The black hole WGC requires that this quantity is negative, and therefore leads to the constraint
\begin{align}
    \label{5dbhwgc}
    \boxed{-2 c_1  -c_2 + 24 c_3 +12 c_4> 0} \, .
\end{align}
As a non-trivial check, we see that this mass shift vanishes for the supersymmetric combination (\ref{SUSYcoeff}) since the extremal Reissner-Nordstrom solution is BPS and cannot receive any mass corrections at fixed charge.

\smallskip\noindent
{\bf \underline{5d Magnetic Black Strings}:}
That some version of the WGC should apply to magnetic charges was pointed out from the very beginning \cite{ArkaniHamed:2006dz}. In a low-energy EFT containing only gravitational and gauge fields, assigning electric vs.\ magnetic charge to a solution reflects an arbitrary choice of electromagnetic duality frame. Thus, the charges should be treated on equal footing. In 4d, dyonic black holes are possible, and the associated continuous family of WGC bounds was analyzed in \cite{Jones:2019nev}. In 5d, in the duality frame with a 2-form field strength, the magnetically charged solutions are black strings.  Together with the bound (\ref{5dbhwgc}) from the electric black hole, we have only a discrete family of bounds. The black string solutions are characterized by mass $M$ and magnetic charge $P$ per unit length, subject to the two-derivative extremality bound 
\begin{equation}
    M\geq \frac{3}{4} |P|\,.
\end{equation}
Extremal black strings saturate this bound and have zero temperature. Including the four-derivative corrections, the corresponding mass shift is  
\begin{align}
    (\Delta M)_{T = 0} = - \frac{3}{40 |P|}( 2 c_1 + 7 c_2 + 24 c_3 + 12 c_4 )\, .
\end{align}
Thus, the constraint implied by the WGC is 
\begin{align}
\label{5Dmag}
    \boxed{ 2 c_1 + 7 c_2 + 24 c_3 + 12 c_4 > 0} \, .
\end{align}
This again vanishes in the supersymmetric case with Wilson coefficients given by (\ref{SUSYcoeff}), as expected.

\section{Dimensional Reduction}

We now ask whether the WGC in a theory obtained by dimensional reduction imposes \emph{new constraints} not visible in the original higher-dimensional theory. To this end, we look for solutions of the theory (\ref{eq:2derLag}) that are locally of the form $\mathds{R}^{3+1}\times S^1$, obtained as the uplift of solutions to an effective 4d KK theory. 

The dimensionally reduced 4d theory is obtained from the reduction Ansatz
\begin{align}
d\hat s_5^2&=e^{\phi/ \sqrt{3}}g_{\mu\nu}dx^\mu dx^\nu+e^{-2 \phi / \sqrt{3} }(dz+\mathcal A)^2,\nn\\
\hat A&=A_\mu dx^\mu+\chi(dz+\mathcal A) \, ,
\end{align}
where $g_{\mu\nu}$ is the 4d metric, the \textit{photon} $A_\mu$ and the \textit{graviphoton} $\mathcal{A}_\mu$ are 4d 1-form gauge fields, while the \textit{dilaton} $\phi$ and the \textit{axion} $\chi$ are a scalar and pseudoscalar respectively. The result of the reduction of the 5d, two-derivative Lagrangian (\ref{eq:2derLag}) is the 4d effective Lagrangian
\begin{align}
        & -16 \pi e^{-1} \mathcal{L}_4 =  \, R \, - \frac{1}{4} e^{-\sqrt{3} \phi } \, G^2   -\frac{1}{2} (\nabla \phi)^2 \nonumber \\
\label{eq:ReducedAction2der}
        &\qquad  -  \frac{1}{4} e^{- \phi / \sqrt{3}} (F + \chi G)^2 - \frac{1}{2} e^{2 \phi/ \sqrt{3}} (\nabla \chi)^2 \\
        & \qquad  +3\lambda\epsilon^{\mu\nu\rho\sigma}\chi\left(F_{\mu\nu}F_{\rho\sigma}+\chi G_{\mu\nu}F_{\rho\sigma}+\frac{1}{3}\chi^2G_{\mu\nu}G_{\rho\sigma}\right)  \, , \nonumber
\end{align}
where $G \equiv d\mathcal A$ and $F \equiv dA$. The electric and magnetic charges 
associated with $G$ will be labelled $Q_0$ and $P_0$ respectively, and the charges associated with $F$ will be labelled by $Q_1$ and $P_1$. 

The 5d Einstein-Maxwell theory with $\lambda^{-1}=12\sqrt3$ represents the bosonic sector of $\mathcal N=2$ supergravity, which reduces to 4d $\mathcal N=2$ supergravity coupled to a vector multiplet.  This can be embedded in the 4d STU model by setting the three STU vector multiplets equal to each other.  In general, this theory admits four-charge black holes which can be obtained from the full set of solutions of the STU model \cite{Chow:2013tia,Chow:2014cca} by identifying the three vector multiplets.  However, we restrict our attention to two-charge solutions with vanishing axion.  While this may not yield the strongest bound on the EFT coefficients, it is nevertheless sufficient to demonstrate sensitivity to topological charge and also has the advantage of being insensitive to the Chern-Simons coupling.

\smallskip\noindent
{\bf \underline{$Q_0$ and $P_0$: Boosted KK Monopoles}:} The 4d theory (\ref{eq:ReducedAction2der}) admits a consistent truncation with vanishing $A_\mu$ and $\chi$.  This truncation is the reduction of pure 5d gravity, and the corresponding solutions are the KK black holes parametrized by mass $M$, angular momentum $J$, and charges $Q_0$ and $P_0$ \cite{Gibbons:1985ac,Rasheed:1995zv,Matos:1996km,Larsen:1999pp}.

\begin{figure*}
\centering\includegraphics[width = \textwidth]{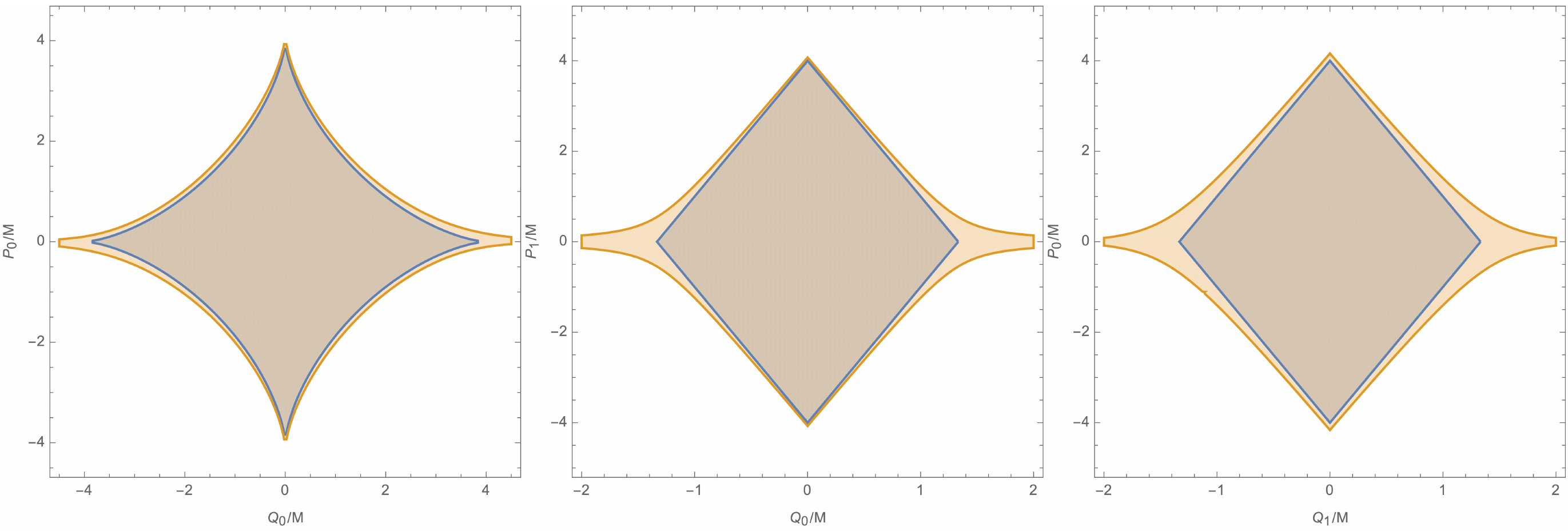}
\caption{
Sub-extremal black hole regions for: \fcolorbox{black}{plotblue}{\rule{0pt}{3pt}\rule{3pt}{0pt}}  two-derivative solutions and \fcolorbox{black}{plotorange}{\rule{0pt}{3pt}\rule{3pt}{0pt}} four-derivative corrected solutions. \textit{Left:} $(Q_0,P_0)$ black hole satisfying WGC bound (\ref{c1pureGR}). \textit{Center:} $(Q_0,P_1)$ black hole satisfying WGC bound (\ref{5Dmag}). \textit{Right:} $(Q_1,P_0)$ black hole satisfying the WGC bounds implied by (\ref{Q1P0shift}), with $c_1=c_2=c_4=c_5=0$ and $c_3>0$.
\label{fig1}}
\end{figure*}

We focus on the static ($J=0$) KK black hole. The mass $M$ and charges $Q_0$, $P_0$ are conveniently expressed in terms of a set of auxiliary parameters $m$, $q$, $p$ via 
\begin{equation}
        M = \frac{p + q}{4}\, , ~ Q_0^2 = \frac{q(q^2 - 4 m^2)}{4(p + q)} \, , ~ P_0^2 = \frac{p(p^2 - 4 m^2)}{4(p + q)} \, ,
\end{equation}
where $0\leq m< \infty$, $2m\leq q<\infty$ and $2m\leq p<\infty$. The temperature and entropy of these solutions are given by
\begin{align}
     T = \frac{m (p + q)}{\pi\sqrt{p q} (2 m + p)(2 m + q)},\qquad S = \frac{m}{2T}.
\end{align}
General solutions are subject to the extremality bound, shown in blue on the left panel of Fig.~\ref{fig1},
\begin{align}
\label{KKboostext}
M \geq \frac{1}{2} \left(Q_0^{2/3}+P_0^{2/3}\right)^{3/2} \, ,
\end{align}
which is saturated at $m=0$. For dyonic ($Q_0$ and $P_0\neq 0$) solutions, the extremal limit has a regular horizon with $S\neq 0$ and $T=0$. However, if any one of the charges vanish, the extremal limit becomes singular at the horizon with $S=0$ and $T\rightarrow \infty$. We conclude that the only physically meaningful extremal limit for purely electric (magnetic) KK black holes is to take $m \to 0$ first, and then to take $q \to 0$ ($p \to 0$). 

The physical interpretation of these solutions in 4d is that they correspond to dyonic black holes with electric charge $Q_0$, magnetic charge $P_0$ and scalar (dilaton) hair. The physical interpretation of the 5d uplifted solution is that they correspond to pure geometry, namely, a Kaluza-Klein monopole boosted on the $S^1$ fiber \cite{Gross:1983hb, Sorkin:1983ns}. The electric charge $Q_0$ is a measure of the boost, while the magnetic charge $P_0$ is the NUT charge of the induced metric on an equal time slice of the 5d spacetime.

A low-energy observer, who cannot resolve the $S^1$, would treat this as any other 4d charged black hole and conclude the necessity of a WGC bound. The four-derivative extremal mass shift is found to be
\begin{align} 
\label{KKdyonicshift}
\begin{split}
    & (\Delta M)_{T=0}= -\frac{c_1}{4 p(p-q)^2}\Big[(p + q)(p - 4q) \\
    & \qquad \qquad \qquad
    + 6 q^2\sqrt{\frac{p + q}{p-q}} \arctanh \left( \sqrt{\frac{p - q}{p + q}} \right)\Big].
\end{split}
\end{align}
The analysis of the stability of these black holes is interesting since the sub-extremal black hole region (\ref{KKboostext}) is \textit{concave} at the two-derivative level. Only the weaker \textit{convex hull condition} \cite{Cheung:2014vva, Jones:2019nev} would be enough to ensure that the four-dimensional KK black holes are kinematically unstable. 
If $c_1<0$ we find that 
the two-derivative black hole region is not contained in the convex hull of the four-derivative corrected black hole region. If $c_1>0$, shown in orange on the left panel of Fig.~\ref{fig1}, 
we find that 
the corrected black hole region completely encloses the uncorrected region. The low-energy observer would therefore obtain the WGC bound
\begin{align}
\label{c1pureGR}
    \boxed{c_1 > 0} \, .
\end{align}
%
The situation is different for an observer who \textit{can} resolve the $S^1$ and knows that this solution uplifts to pure geometry in 5d.  In this case, since the solution is uncharged under the primordial 5d gauge field, they would conclude that the WGC imposes no such constraint. We therefore have a dilemma in how to apply the WGC. Either: \textit{(i)} the WGC only applies to primordial gauge fields and not those arising from dimensional reduction; this would render it useless as a Swampland criterion since this distinction is invisible to a low-energy observer, or \textit{(ii)} the WGC applies independently of the physical origin of the gauge field in the UV completion. The second possibility is the more intriguing and conceptually coherent, and will be the position advocated for in this letter.

The constraint (\ref{c1pureGR}) on the sign of the Gauss-Bonnet coefficient in $d>4$ has previously been argued from unitarity assuming a tree-level UV completion \cite{Cheung:2016wjt}, and from entropy bounds on large black holes in AdS \cite{Cremonini:2019wdk}. \\
\\
{\bf \underline{$Q_0$ and $P_1$: Boosted Magnetic Black Strings}:} We now consider 4d black holes that uplift to 5d charged black hole solutions. In particular, we take solutions with electric charge $Q_0$ under the graviphoton, and magnetic charge $P_1$ under the photon. The extremal limit of these solutions lies on the BPS branch of STU black holes; they remain solutions of the general 4d theory (\ref{eq:ReducedAction2der}), as they are insensitive to the Chern-Simons coupling. 
 General solutions are subject to the extremality bound shown in blue on the center panel of Fig.~\ref{fig1},
 \begin{align}
    M \geq \frac{1}{4} \left(|Q_0| + 3|P_1| \right) \, ,
\end{align}
which is saturated, for general $Q_0$ and $P_1$, at $T=0$.
The four-derivative extremal mass shift is found to be 
\begin{align}
(\Delta M)_{T=0} = -\frac{3 }{40 |P_1|}\left(2 c_1+7c_2+24c_3+12c_4\right).
\end{align}
Note that this is the same combination of EFT coefficients that appeared for the 5d magnetic black string. Thus, we arrive at the same condition as in (\ref{5Dmag}), 
displayed in the center panel of Fig.~\ref{fig1}. 
We conclude that in this case turning on $Q_0$ has no effect on the WGC bound, unlike for the dyonic black hole result (\ref{KKdyonicshift}) which is sensitive to both charges. This is an interesting feature that we would like to understand better.\\
\\
{\bf \underline{$Q_1$ and $P_0$: Electrically Charged KK Monopoles}:} We can immediately obtain a black hole with the opposite set of charges by applying 4d electric/magnetic duality to the $(Q_0,P_1)$ solution. The resulting 4d black hole has electric charge $Q_1$ under the photon and magnetic charge $P_0$ under the graviphoton. When lifted to 5d, $Q_1$ corresponds to electric charge while, as discussed above, $P_0$ corresponds to the NUT charge of an equal time slice of the 5d spacetime. General solutions are subject to the extremality bound shown in blue on the right panel of Fig.~\ref{fig1},
\begin{align}
    M \geq \frac{1}{4} \left(|P_0| + 3|Q_1| \right) \, ,
\end{align}
which is saturated for general $Q_1$ and $P_0$ at $T=0$.

The four-derivative extremal mass shift is given by
\begin{align}
\label{Q1P0shift}
    & (\Delta M)_{T = 0} = \frac{ 1}{32 (P_0 - Q_1)^5} \Bigg[ c_1  \biggl( 24 Q_1^2 (Q_1^2 - 8P_0^2) \log \left( \frac{P_0}{Q_1} \right)\nn \\
    &\qquad  - 2 (P_0 - Q_1)(5 P_0^3 - 43 P_0^2 Q_1 - 79 P_0 Q_1^2 + 33 Q_1^3)   \biggr) \nn \\
    & \qquad + (c_2 - 24 c_3 - 12 c_4)  \biggl( 12 Q_1^4 \log \left( \frac{P_0}{Q_1} \right) + (P_0 - Q_1)\nn  \\
    & \qquad \times (3 P_0^3 - 13 P_0^2 Q_1 + 23 P_0 Q_1^2 - 25 Q_1^3)\biggr)\nn\\
    &\qquad+32\sqrt3c_5\biggl(12P_0^2Q_1^2\log\left(\fft{P_0}{Q_1}\right)+(P_0^2-Q_1^2)\nn\\
    &\qquad\times(P_0^2-8P_0Q_1+Q_1^2)\biggr)
    \Bigg]\, ,
\end{align}
where all charges should be understood to be their absolute values.  Note that, when deriving the mass shift from the free energy, some care is needed in defining the gauge-invariant regulated on-shell action for the $\hat A\wedge\hat R\wedge\hat R$ term corresponding to the Wilson coefficient $c_5$.  As a check, we verify that this mass shift vanishes when the four-derivative corrections preserve $\mathcal{N}=2$ supersymmetry (\ref{SUSYcoeff}).

The bounds on the coefficients imposed by the WGC depend on the values of $P_0$ and $Q_1$. However, the strongest bounds come from the most extreme cases; when $P_0 \rightarrow 0$ we find exactly the same bound we derived for electric black holes in 5d, given by (\ref{5dbhwgc}).
On the other hand, when $Q_1\rightarrow 0$ we find
\begin{align}
\label{Q20}
\boxed{\frac{10}{3} c_1 - c_2 + 24 c_3 + 12 c_4-\fft{32}{\sqrt3}c_5 > 0} \, .
\end{align}
This bound is new and independent of the WGC bounds implied by the 5d electric black hole (\ref{5dbhwgc}) and the 5d magnetic black string (\ref{5Dmag}). Thus, the WGC bounds implied by the extremal solutions of the $S^1$ reduction are \textit{strictly stronger} than those implied by the extremal, asymptotically flat, solutions. 

To be conceptually consistent, it ought to be possible to state the rationale behind these bounds equally well in both 4d and 5d language. 
To discover (\ref{Q20}) without dimensional reduction (as well as the family of bounds implied by (\ref{Q1P0shift}) that interpolate between (\ref{5dbhwgc}) and (\ref{Q20})), we would have had to consider the WGC for 5d solutions with topological charges, such as the NUT charge of the KK monopole considered here. 

\section{Conclusions}

In this letter, we have argued that the WGC should be extended to include topological charges associated with the global structure of spacetime. After KK reduction these charges appear to a low-energy observer as ordinary magnetic gauge charges, so the familiar logic supporting the WGC should apply. 

In combining bounds derived from different backgrounds, we have implicitly assumed that to determine if an EFT lies in the landscape, we should consider all possible solutions including lower dimensional compactifications.
However, 
it may be instead that the swampland is defined by a low-energy action 
\textit{plus} information about the superselection sector, such as the boundary topology, with the WGC satisfied in some sectors but not others.
%
%
%
Determining how to interpret such topological charges and 
whether the WGC should generically apply to them is therefore an important step to understand exactly what defines 
the swampland.

A complementary approach to shed light on these issues 
is to connect these strengthened bounds to unitarity/causality of scattering amplitudes \cite{Bellazzini:2019xts, Chen:2019qvr, Hamada:2018dde, Cheung:2016wjt}. 
The latter do not appear to coincide with 
the WGC bounds in general \cite{Andriolo:2020lul, Loges:2020trf}. Thus, understanding the minimal set of additional UV assumptions 
needed to imply the WGC is an important open problem. It would also be interesting to apply the methods of \cite{NewSpin} to the solutions we present, as many of them have three-dimensional near-horizon geometries. Finally, we leave for future work the problem of extending the analysis presented in this letter to general dimensions, including calculating the full set of WGC bounds obtained from extremal branes with combinations of topological and $p$-form charges, as described in \cite{Hull:1997kt}.

\section*{Acknowledgments}

We would like to thank Gary Shiu for comments on the draft. This work was supported in part by the U.S.~Department of Energy under grant DE-SC0007859. 
SC is supported in part by the National Science Foundation grant PHY-1915038. CRTJ and BM were supported by Leinweber research fellowships. YT is supported by Lehigh University's Lee Fellowship.

\bibliography{cite.bib}


\section{Appendix: Five- and four-dimensional gravitational solutions}

In this supplemental material we summarize the black string and black hole solutions used in the body of the letter.  These solutions have been well established in the literature, and additional details can be found in the references.

\subsection{Five-dimensional solutions}

Five-dimensional Einstein-Maxwell theory, with the Lagrangian
\begin{equation}
    -16\pi\hat e^{-1}\mathcal L_5=\hat R-\fft14\hat F^2,
\label{eq:5dem}
\end{equation}
admits electrically charged black hole and magnetically charged black string solutions.  Here we have suppressed the Chern-Simons term as the static solutions we consider are insensitive to it.

\smallskip\noindent
{\bf \underline{5D Electric Black Holes}:}
The electric black hole is just the five-dimensional Reissner-Nordstrom solution
\begin{equation}
\begin{split}
    d\hat s^2 &= -f\, dt^2 + \fft{dr^2}f + r^2 d \Omega_{S^3}^2\\
    \hat A &= \sqrt{3}\frac{ Q}{r^2} dt \, ,
\end{split}
\end{equation}
where
\begin{equation}
    f(r) = 1 - \frac{2 M}{r^2} + \frac{Q^2}{r^4} \, .
\end{equation}
This solution is characterized by mass $M$ and charge $Q$.  Note in particular that our normalization of charge in geometric units pulls out a $\sqrt3$ factor in the gauge potential.

The Reissner-Nordstrom black hole has outer and inner horizons located at
\begin{align}
    r_\pm = (M \pm\sqrt{M^2 - Q^2})^{1/2} \, ,
\end{align}
and a temperature given by
\begin{align}
    T = \frac{Q^2 - M^2 - M \sqrt{M^2 - Q^2} }{\pi (M + \sqrt{M^2 - Q^2})^{5/2}} \, .
\end{align}
The two horizons coincide and the temperature vanishes at extremality, which occurs when $M = Q$. 

\smallskip\noindent
{\bf \underline{5D Magnetic Black Strings}:}
The magnetic black string can be thought of as part of a family of black $p$-brane solutions in diverse dimensions \cite{Horowitz:1991cd,Duff:1993ye,Lu:1993vt,Lu:1995cs,Duff:1996hp}.  The 5d black string takes the form
\begin{equation}
\begin{split}
    d\hat s^2 &= \frac{1}{H}(-f\,dt^2+dz^2 ) + H^2\left(\frac{dr^2}{f} + r^2 d \Omega_{S^2}^2\right) \, , \\
    \hat A &= \sqrt3 P\cos \theta \  d\varphi \, ,
\end{split}
\end{equation}
where
\begin{equation}
    H = 1 + \frac{\mu\sinh^2\beta}{r},\quad  f = 1 -\frac{\mu}{r} \, .
\end{equation}
The magnetic charge and mass per unit length are given in terms of the parameters $\mu$ and $\beta$ as
\begin{equation}
    P=\mu\sinh\beta\cosh\beta\,,\qquad M=\fft14\mu(3\sinh^2\beta+2)\,.
\end{equation}

In the above parametrization, the outer horizon is located at $\mu$ and the inner horizon is at $0$.  The temperature and entropy per unit length are
\begin{equation}
    T=\fft1{4\pi\mu\cosh^3\beta},\qquad S=\pi\mu^2\cosh^3\beta\,.
\end{equation}
Note that the extremal limit is obtained by taking $\mu\to0$ while holding the charge $P$ fixed.  The temperature vanishes in this limit while the extremal mass becomes $M=(3/4)P$.

\subsection{Four-dimensional solutions}

The five-dimensional Einstein-Maxwell Lagrangian, (\ref{eq:5dem}), reduces on a circle to four-dimensional gravity coupled to two gauge fields $A$ and $\mathcal A$ and two scalars $\phi$ and $\chi$.  The four-dimensional solutions considered in the body of the letter all have vanishing axion $\chi$ and can be obtained from the Lagrangian
\begin{equation}
    -16\pi e^{-1}\mathcal L=R-\fft12\partial\phi^2-\fft14e^{-\sqrt3\phi}G^2-\fft14e^{-\phi/\sqrt3}F^2\,,
\label{eq:truncL}
\end{equation}
where $F=dA$ and $G=d\mathcal A$.  Note that this Lagrangian is not a consistent truncation as we need to impose the constraint $F_{\mu\nu}G^{\mu\nu}=0$ for the solutions to remain valid in the absence of the axion. A consistent truncation to pure 4d Einstein-Maxwell theory can be obtained by setting $\phi=0$ and $F=\sqrt3*G$, provided we include the 5d Chern-Simons term in (\ref{eq:5dem}) as in the bosonic sector of 5d $\mathcal N=2$ supergravity.

The 4d black holes can carry charges $Q_0$ and $P_0$ of the graviphoton $\mathcal A$ and charges $Q_1$ and $P_1$ of the photon $A$.  The $F_{\mu\nu}G^{\mu\nu}=0$ constraint then corresponds to the requirement
\begin{equation}
    Q_0Q_1=P_0P_1\,.
\label{eq:QQPPcons}
\end{equation}
All the solutions we consider trivially satisfy this constraint.

\smallskip
\noindent
{\bf \underline{4D, $Q_0$ and $P_0$ charge}:}
By setting the photon $A$ to zero, the Lagrangian (\ref{eq:truncL}) reduces to the original Kaluza-Klein reduction of 5d Einstein gravity.  This is in fact a consistent truncation as the $F_{\mu\nu}G^{\mu\nu}=0$ constraint is now trivially satisfied.  Static Kaluza-Klein black holes carrying $Q_0$ and $P_0$ charges were constructed in \cite{Gibbons:1985ac}, and rotating ones in \cite{Rasheed:1995zv,Matos:1996km,Larsen:1999pp} (see also \cite{Horowitz:2011cq}).  Following the notation of \cite{Larsen:1999pp}, the solution is given by
\begin{align}
    & ds^2 \, = \, - \frac{\Delta}{\sqrt{H_1 H_2}} \, dt^2 + \frac{\sqrt{H_1 H_2}}{\Delta} \, dr^2 + \sqrt{H_1 H_2} \, d \Omega_{S^2}^2 \, ,\nn \\
    & \mathcal{A} = - \, Q_0 \,(2r+ p -  2m) \, H^{-1}_2 \, dt - 2 P_0 \, \cos \theta \, d \varphi  \, ,\nn \\
    & e^{-2\phi/\sqrt{3}} = \frac{H_2}{H_1} \, ,
\end{align}
where 
\begin{align}
    \begin{split}
        & \Delta = r^2 - 2 m r \, , \\
        & H_1 = r^2 + r(p - 2 m) + \frac{p (p - 2m)(q - 2 m)}{2 (p + q)}\, , \\
        & H_2 = r^2 + r(q - 2 m) + \frac{q (p - 2m)(q - 2 m)}{2 (p + q)}\, . \\
    \end{split}
\end{align}
Here $m\ge0$ is a non-extremality parameter and $p\ge2m$ and $q\ge2m$ are related to the physical charges $Q_0$, $P_0$ by
\begin{align}
    \begin{split}
        & Q_0^2 = \frac{q(q^2 - 4 m^2)}{4(p + q)} \, , \qquad P_0^2 = \frac{p(p^2 - 4 m^2)}{4(p + q)}\,.
    \end{split}
\end{align}
Purely electric black holes are given by setting $p=2m$ and purely magnetic ones by $q=2m$.

The mass of the black hole is given by
\begin{equation}
    M=\fft{p+q}4\,,
\end{equation}
while the temperature and entropy are
\begin{equation}
    T = \frac{m (p + q)}{\pi\sqrt{p q} (2 m + p)(2 m + q)}, \qquad S = \frac{m}{2T}.
\end{equation}
The outer and inner horizons are located at $r=2m$ and $0$, and extremality is obtained in the limit $m\to0$.  However, there is a subtlety in how to take the limit when either one of the charges vanish.  Consider, for example, a purely electric black hole with vanishing magnetic charge $P_0$ obtained by setting $p=2m$.  In this case, the temperature becomes
\begin{align}
    T = \frac{1}{4 \pi \sqrt{2 q m}}\,,
\end{align}
which diverges in the extremal limit when $m \to 0$.  In order to obtain zero temperature at extremality, we have to instead take $m\to0$ first before taking $p\to0$.  A similar procedure can be used for the purely magnetic solutions.  Dyonic black holes always have zero temperature in the extremal limit.

Finally, we note that these 4d solutions may be uplifted to pure gravity in 5d, where they take the form
\begin{align}
    ds_5^2 = - \frac{\Delta}{H_2} dt^2 +  \frac{H_1}{\Delta} dr^2 + H_1 d \Omega_{S^2}^2+\frac{H_2}{H_1} (dz + \mathcal{A})^2\,.
\end{align}
In 5d, we see that the electric charge $Q_0$ leads to a $dt dz$ component, so its value is related to a boost of the solution along the $z$ direction. The magnetic charge $P_0$ on the other hand gives $d \varphi dz$ components. This may be interpreted as a NUT charge in an equal time slice of the 5d solution. In the limit where $Q_0 \to 0$, this solution factors into time and a Euclidean Taub-NUT solution \cite{Horowitz:2011cq}.

\smallskip
\noindent
{\bf \underline{4D, $Q_0$ and $P_1$ charge}:}
The constraint (\ref{eq:QQPPcons}) precludes the possibility of turning on all four charges without at the same time sourcing the axion $\chi$.  Motivated by supergravity black holes, and in particular the BPS branch of STU black holes, we consider solutions with either non-vanishing $(Q_0,P_1)$ or non-vanishing $(P_0,Q_1)$.

The 4d STU model consists of 4d $\mathcal N=2$ supergravity coupled to three vector multiplets and can be obtained from the dimensional reduction of the 5d STU model which has 5d $\mathcal N=2$ supergravity coupled to two vector multiplets.  To make connection with 5d Einstein-Maxwell, (\ref{eq:5dem}), note that the 5d STU model admits a truncation to pure 5d $\mathcal N=2$ supergravity by setting the three $U(1)$ fields equal to each other.  In this case, the bosonic Lagrangian is given by
\begin{equation}
    -16\pi\hat e^{-1}\mathcal L_5=\hat R-\fft34\hat F^2+\hat F\wedge\hat F\wedge\hat A\,,
\label{eq:5dN=2}
\end{equation}
where the factor of three in the Maxwell term arises from identifying the three $U(1)$ fields but can be removed by the scaling $\hat A\to\hat A/\sqrt3$.

Reduction of (\ref{eq:5dN=2}) on a circle yields 4d $\mathcal N=2$ supergravity coupled to a single vector multiplet.  This corresponds to identifying the three vector multiplets of the 4d STU model, yielding what may be denoted the S$^3$ model.  The bosonic Lagrangian corresponds to equation (11) in the letter with  $\lambda^{-1}=12\sqrt3$.  More precisely, taking the STU model to be defined by the prepotential $F=X^1X^2X^3/X^0$, the identification of the fields in equation (11) in the letter is
\begin{equation}
\begin{split}
    A^0=\mathcal A,\qquad A^1&=A^2=A^3=\fft1{\sqrt3}A\,,\\
    z^1&=z^2=z^3=-\fft1{\sqrt3}\chi-ie^{-\phi/{\sqrt3}}\,,
\end{split}
\end{equation}
where $z^i=X^i/X^0$ ($i=1,2,3$) and the $\sqrt3$ factors are inserted to obtain canonical normalization of the identified fields.

The 4d STU model admits four-charge dyonic BPS black holes as well as their non-extremal generalizaztions \cite{Cvetic:1995uj,Cvetic:1995dn} (see also \cite{Chow:2013tia,Chow:2014cca} and references therein for the most general rotating non-extremal solution).  To make contact with the two $U(1)$ system, (\ref{eq:truncL}), we set $Q^1=Q^2=Q^3$ as well as $P^1=P^2=P^3$.  Assuming a vanishing axion then allows a BPS solution with either $(Q_0,P_1)$ charges or $(P_0,Q_1)$ charges.

The solution with an electric charge $Q_0$ for the graviphoton $\mathcal{A}$ and a magnetic charge $P_1$ for the photon $A$ takes the form:
\begin{align}
    & ds^2 = - (H_0H_1^3)^{-1/2}f\, dt^2 + (H_0H_1^3)^{1/2}\left(\fft{dr^2}f+r^2d \Omega_{S^2}^2\right),\nn\\
    &\mathcal{A} = \left(1-\fft1{H_0}\right)\coth\beta_0dt,\quad
    e^{-2 \phi/\sqrt3} = \fft{H_0}{H_1}\,,\nn\\
    &A =\sqrt3\mu\sinh\beta_1\cosh\beta_1 \cos\theta d\varphi\,,
\end{align}
with
\begin{equation}
    H_0 =1+\fft{\mu\sinh^2\beta_0}r,\quad
    H_1 =1+\frac{\mu\sinh^2\beta_1}{r}, \quad
    f = 1-\frac{\mu}{r}\, .
\label{eq:H0H1f}
\end{equation}
The mass and charges of this solution are given by
\begin{align}
\begin{split}
    &M = \frac{\mu}{8} \left( \cosh 2 \beta_0 + 3\cosh 2 \beta_1 \right) \, , \\
    &Q_0 = \mu \sinh\beta_0 \cosh\beta_0 \, , \\
    &P_1 = \mu \sinh\beta_1 \cosh\beta_1 \, .
\end{split}
\label{eq:MQP}
\end{align}

The horizon is located at $r=\mu$, and the corresponding Hawking temperature and entropy are given by 
\begin{align}
\begin{split}
    T&=\fft1{4\pi\mu\cosh\beta_0\cosh^3\beta_1}\,,\\
    S&=\pi\mu^2\cosh\beta_0\cosh^3\beta_1\,.
\end{split}
\label{eq:TandS}
\end{align}
The extremal limit is taken by letting $\mu\to0$ and $\beta_i\to\infty$ with $Q_0$ and $P_1$ fixed.  In this limit, we find
\begin{equation}
    M=\fft14(|Q_0|+3|P_1|),\quad S=\pi\sqrt{|Q_0P_1^3|},\quad T=0\,.
\label{eq:MSTBPS}
\end{equation}
The black hole is BPS when the signs of the charges are chosen appropriately; otherwise it's extremal but non-BPS.

\smallskip
\noindent
{\bf \underline{4D, $Q_1$ and $P_0$ charge}:}
The solution with non-zero $Q_1$ and $P_0$ can be obtained by taking the overall electric/magnetic dual of the $(Q_0,P_1)$ solution given above.  To be explicit, it takes the form
\begin{align}
    & ds^2=-(H_0H_1^3)^{-1/2}f\,dt^2+(H_0H_1^3)^{1/2}\left(\fft{dr^2}f+r^2d\Omega_{S^2}^2\right),\nn\\
    & \mathcal A=\mu\sinh\beta_0\cosh\beta_0\cos\theta d\varphi,\qquad
    e^{-2 \phi/\sqrt3}=\frac{H_1}{H_0}\,,\nn\\
    & A=\sqrt3\left(1-\fft1{H_1}\right)\coth\beta_1dt\,,
\end{align}
with the functions $H_0$, $H_1$ and $f$ defined in (\ref{eq:H0H1f}).  Note that duality flips the sign of the dilaton, which has the implication that the lifted 5d metrics of the $(Q_0,P_1)$ and $(P_0,Q_1)$ systems are not identical.

From the 4d point of view, the conserved charges and thermodynamic quantities are given by (\ref{eq:MQP}), (\ref{eq:TandS}) and (\ref{eq:MSTBPS}) with the duality replacement $Q_0\to P_0$ and $P_1\to Q_1$.  In particular, in the extremal limit, we find
\begin{equation}
    M=\fft14(|P_0|+3|Q_1|),\quad S=\pi\sqrt{|P_0Q_1^3|},\quad T=0\,.
\end{equation}
%


\end{document}